\documentclass[twocolumn,showpacs,amsmath,pre]{revtex4}
\usepackage{graphicx}
\usepackage{amsmath}
\usepackage{amssymb}
\usepackage{natbib}
\usepackage{subfigure}
\usepackage{color} 

\begin{document}

\title{Vibrations in jammed solids: Beyond linear response}

\author{Thibault Bertrand$^{1}$}
\author{Carl F. Schreck$^{1,2}$}
\author{Corey S. O'Hern$^{1,2,3}$}
\author{Mark D. Shattuck$^{4,1}$}

\affiliation{$^{1}$Department of Mechanical Engineering \& Materials Science, Yale University, New Haven, Connecticut 06520-8260, USA}

\affiliation{$^{2}$Department of Physics, Yale University, New Haven,
Connecticut 06520-8120, USA}

\affiliation{$^{3}$Department of Applied Physics, Yale University, New Haven, Connecticut 06520-8120, USA}

\affiliation{$^{4}$Benjamin Levich Institute and Physics Department, The City College of the City University of New York, New York, New York 10031, USA}

\begin{abstract}
We propose a `phase diagram' for particulate systems that interact via
purely repulsive contact forces, such as granular media and colloidal
suspensions.  We identify and characterize two distinct classes of
behavior as a function of the input kinetic energy per degree of
freedom $T_0$ and packing fraction deviation above and below jamming
onset $\Delta \phi=\phi - \phi_J$ using numerical simulations of
purely repulsive frictionless disks.  Iso-coordinated solids (ICS)
only occur above jamming for $\Delta \phi > \Delta \phi_c(T_0)$; they
possess average coordination number equal to the isostatic value
($\langle z\rangle = z_{\rm iso}$) required for mechanically stable
packings.  ICS display harmonic vibrational response, where the
density of vibrational modes from the Fourier transform of the
velocity autocorrelation function is a set of sharp peaks at
eigenfrequencies $\omega_k^d$ of the dynamical matrix evaluated at
$T_0=0$.  Hypo-coordinated solids (HCS) occur both above and below
jamming onset within the region defined by $\Delta \phi > \Delta
\phi^*_-(T_0)$, $\Delta \phi < \Delta \phi^*_+(T_0)$, and $\Delta \phi
> \Delta \phi_{cb}(T_0)$. In this region, the network of interparticle
contacts fluctuates with $\langle z\rangle \approx z_{\rm iso}/2$, but
cage-breaking particle rearrangements do not occur.  The HCS
vibrational response is nonharmonic, {\it i.e.}  the density of
vibrational modes $D(\omega)$ is not a collection of sharp peaks at
$\omega_k^d$, and its precise form depends on the measurement method.
For $\Delta \phi > \Delta \phi_{cb}(T_0)$ and $\Delta \phi < \Delta
\phi^*_{-}(T_0)$, the system behaves as a hard-particle liquid.

\end{abstract}


\maketitle

\begin{figure}
\includegraphics[scale=0.31]{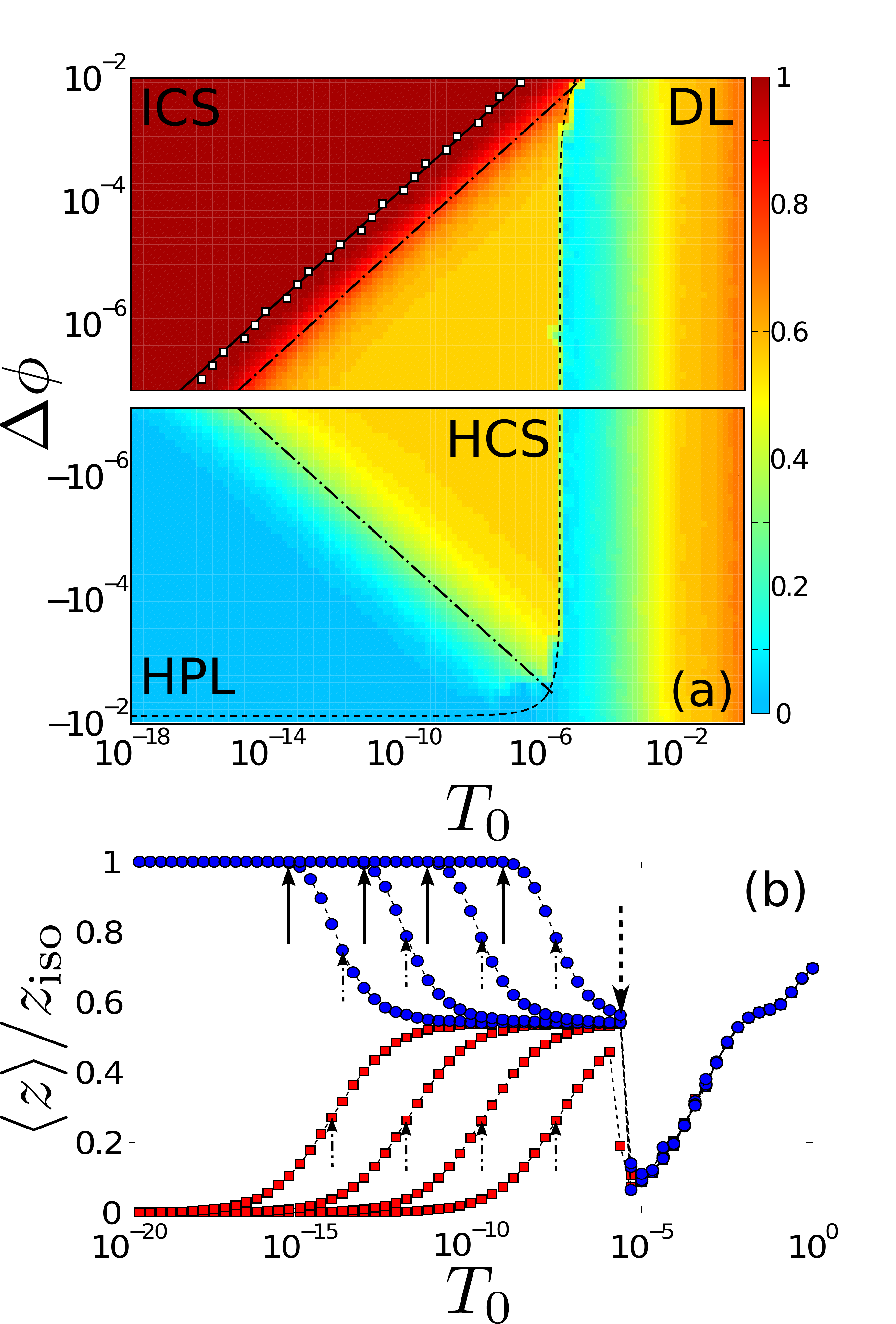}
\caption{(Color online) (a) `Phase diagram' for the vibrational
response of MS packings versus $\Delta \phi$ and $T_0$ illustrated for
$N=10$. The shading gives the time-averaged contact number $\langle
z\rangle/z_{\rm iso}$. For $0<\Delta \phi(T_0) < \Delta \phi_c(T_0)$
(data and scaling curve given by open squares and solid line,
respectively), the contact network for the `ICS' does not change from
that at $T_0=0$ and the vibrational response is harmonic.  The
midpoints $T_0$ at which $\langle z\rangle/z_{\rm iso}$ crosses over
from $1$ to $0.5$ and from $0$ to $\approx 0.5$, which define $\Delta
\phi^*_{\pm}(T_0)$, are indicated by dot-dashed lines (with slope $\pm
0.5$).  In the `HCS' and `HPL' regions, the contact network fluctuates
with $\langle z\rangle/z_{\rm iso} \approx 0.5$ and $0$, respectively,
but there are no particle rearrangements as in the `DL' regime with
$\Delta \phi < \Delta \phi_{cb}(T_0)$ (dashed line). (b) $\langle
z\rangle/z_{\rm iso}$ versus $T_0$ for $\Delta \phi = \pm 10^{-7}$,
$\pm 10^{-6}$, $\pm 2\times 10^{-5}$, and $\pm 2 \times 10^{-4}$
(circles and squares) from left to right. The solid, dot-dashed, and
dashed arrows indicate $\Delta \phi_c(T_0)$, $\Delta
\phi^*_{\pm}(T_0)$, and $\Delta \phi_{cb}(T_0)$, respectively, in
(a).}
\label{phase_diagram}
\end{figure}

The vibrational response of conventional solids, such as metals,
ceramics, and minerals, can be described by linear response at
sufficiently low temperatures compared to the melting
points~\cite{montroll}.  Nonlinearities stemming from weak structural
disorder and the shape of the interaction potential explored at low
temperatures can be treated as small perturbations to the description
of the harmonic solid~\cite{pathak}.  Particulate systems, such as
granular media~\cite{dauchot} and colloids~\cite{yodh}, can also exist
in solid-like states in the limit of weak driving or thermal
fluctuations.  However, in contrast to molecular-scale
solids, the interactions in many particulate solids
are purely repulsive and vanish when particles come out of contact.
We have shown previously~\cite{schreck} that even small changes
in the contact network in purely repulsive particle-based solids give
rise to strong nonlinearities in the measured vibrational response
that are not found in conventional solids.

In spite of these nonlinearities, a major emphasis of the jamming
literature in the past decade has been to invoke linear response of
static packings to provide insight into the structural relaxation of
dense liquids near the glass transition~\cite{review1,review2}.
However, one of the most obvious and important questions has been left
unanswered: what is the measured response of static packings near
jamming onset when they are subjected to vibrations? In this Letter,
we do not rely on linear response to infer vibrational
behavior. Instead, we measure directly the vibrations of model
particulate systems as a function of $\Delta \phi=\phi-\phi_J$ above
and below jamming and input kinetic energy $T_0$.

We identify two distinct classes of behavior in the $\Delta \phi$ and
$T_0$ plane near jamming as shown in Fig.~\ref{phase_diagram}:
iso- and hypo-coordinated solids (ICS and HCS), which
are distinguished by the time-averaged contact number $\langle z
\rangle$ and density of vibrational modes $D(\omega)$.  We focus on
the regime where there are no cage-breaking particle rearrangements.
For the ICS, with $\Delta \phi > \Delta \phi_c(T_0) > 0$, the contact
network does not change from that at $T_0=0$, and the vibrational
response is harmonic with strong peaks in the Fourier transform of the
velocity autocorrelation function at the dynamical matrix
eigenfrequencies.  HCS occur both above and below
$\phi_J$~\cite{wyart} in the region defined by $\Delta \phi > \Delta
\phi^*_-(T_0)$, $\Delta \phi < \Delta \phi^*_+(T_0)$, and $\Delta \phi
> \Delta \phi_{cb}(T_0)$. In the HCS, the network of interparticle
contacts fluctuates with an average contact number at approximately
half of the isostatic value ($\langle z\rangle/z_{\rm iso} \approx
0.5$), the vibrational response is strongly nonharmonic, and the form
of $D(\omega)$ depends on the measurement method. In the regime
$\Delta \phi > \Delta \phi_{cb}(T_0)$ and $\Delta \phi < \Delta
\phi^*_{-}(T_0)$, $\langle z\rangle/z_{\rm iso} \sim 0$, and
$D(\omega)$ resembles that for hard-particle liquids (HPL).

{\it Model and Simulations}
We measure the vibrational response of mechanically stable (MS)
packings of $N$ bidisperse frictionless disks with mass $m$ that
interact via the pairwise purely repulsive potential
\begin{equation} 
\label{interaction} 
V(r_{ij}) = \frac{\epsilon}{2} \left( 1-\frac{r_{ij}}{\sigma_{ij}}
\right)^{2} \Theta\left( 1-\frac{r_{ij}}{\sigma_{ij}} \right),
\end{equation} 
where $r_{ij}$ is the separation between disk centers, $\sigma_{ij} =
(\sigma_i + \sigma_j)/2$ is the average disk diameter, $\epsilon$ is
the energy scale of the repulsive interaction, and $\Theta(x)$ is the
Heaviside step function.  The bidisperse mixtures contained half large
and half small disks by number with diameter ratio
$r=\sigma_2/\sigma_1=1.4$.  We focus on the $\Delta \phi \rightarrow
0$ limit for which the MS packings possess the isostatic number of
interparticle contacts $N^{\rm iso}_c = 2N - 1$, where $N =N'-N_r$ and $N$ is
the number of particles after $N_r$ rattler particles with fewer 
than $3$ contacts have been removed.

We generate MS packings at $\Delta \phi_0 = 10^{-8}$ using
the successive compression and decompression protocol described
previously~\cite{gao} for system sizes in the range $N=10$ to $512$.
Each of these packings were then decompressed or overcompressed in a
single step to a given $\Delta \phi$ in the range $-10^{-2} \le \Delta
\phi \le 10^{-2}$ followed by conjugate gradient energy minimization
to the configuration ${\vec R}^0 =
\{x^0_1,y^0_1,\ldots,x^0_N,y^0_N\}$.
 
We perturbed each system at ${\vec R}^0$ by exciting equal
kinetic energy in each mode.  We selected the initial particle
velocities ${\vec v} = \{v_{xi},v_{yi},\dots,v_{xN},v_{yN}\}$
according to
\begin{equation}
v_n = \delta \sum_{k=1}^{2N-2} e^k_{n},  
\end{equation} 
where ${\hat e}^k$ are the $2N-2$ eigenvectors (corresponding to the
nonzero eigenvalues) of the dynamical matrix~\cite{tanguy} evaluated
at ${\vec R}^0$, $({\hat e}^k)^2=1$, and $\delta$ is chosen so that
the normalized kinetic energy $T_0 = {\epsilon}^{-1}\sum_i \frac{1}{2}
mv_i^2/(2N-2)$ is in the range $10^{-20} \le T_0 \le 10^{-1}$. We then
integrated Newton's equations of motion at constant total energy and
area in a square box with side length $L=1$ using the velocity Verlet
algorithm with time step $\Delta t= 1/(400 \pi) \sigma_1
\sqrt{m/\epsilon}$.  We first ran the constant energy simulations for
$10^3$ oscillations of the lowest dynamical matrix eigenfrequency
$\omega_1^d$ and then quantified fluctuations in the particle
positions over the next $10^3$ periods.

\begin{figure}
\includegraphics[scale=0.21]{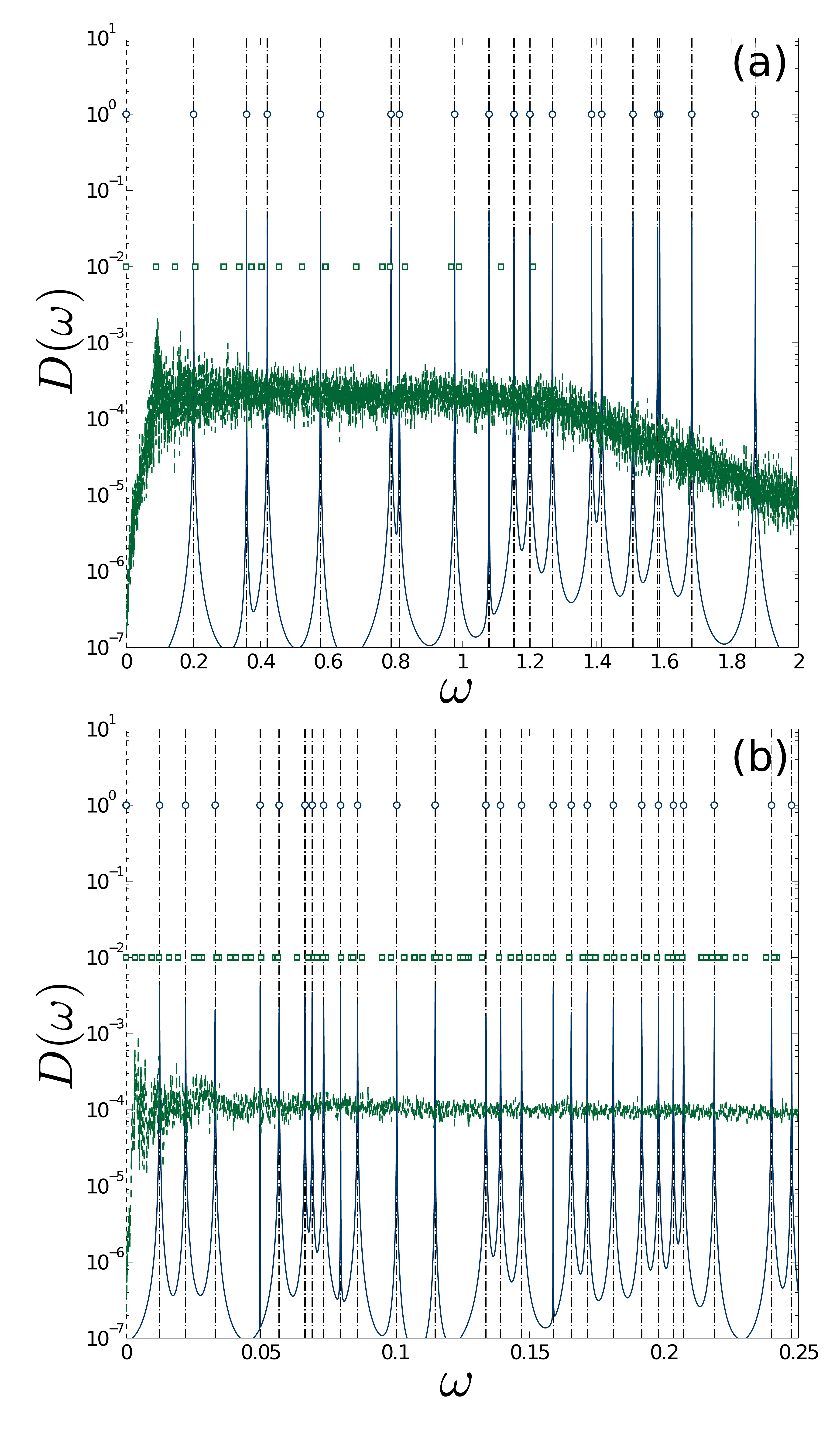}
\vspace{-0.1in}
\caption{(Color online) Comparison of portions of the density of vibrational
frequencies $D(\omega)$ from the Fourier transform of the velocity
autocorrelation function (lines) and associated with the dynamical
(vertical dot-dashed lines) and displacement correlation matrices
(symbols) for the ICS at $\Delta \phi = 2 \times 10^{-7}$ and $T_0 = 4
\times 10^{-19}$ (blue solid lines and circles) and HCS at $\Delta
\phi = 2 \times 10^{-7}$ and $T_0 = 2 \times 10^{-11}$ (green dashed
lines and squares) for (a) $N=10$ and (b) $128$.}
\vspace{-0.1in}
\label{comparison}
\end{figure}

In the harmonic regime, the time-dependent particle positions
are described by 
\begin{equation}
\label{positions}
R_n(t) = R^0_n + \sum_{k=1}^{2N-2} {\widetilde R}_k e^k_{n} \cos(\omega_k
t+\psi_k),
\end{equation}
where ${\widetilde R}_k$ are the time-independent amplitudes of the
normal modes ${\hat e}_k$ with eigenfrequency $\omega^d_k$ from the
dynamical matrix.  We employed two additional methods to measure the
vibrational response as a function of $\Delta \phi$ and
$T_0$.  We calculated the Fourier transform of the normalized velocity
autocorrelation function to quantify the density of vibrational
modes~\cite{berthier}
\begin{equation}
\label{vacf}
D(\omega^v) = \int_0^{\infty} dt \frac{ \langle {\vec v}(t_0+t) \cdot
{\vec v}(t_0) \rangle}{\langle {\vec v}(t_0) \cdot {\vec v}(t_0)
\rangle} e^{i\omega_v t},
\end{equation}
where $\langle .\rangle$ indicate averages over all particles and time
origins $t_0$.  We also measured the eigenvalue spectrum of $S=V
C^{-1}$ (which equals the dynamical matrix $M$ provided
Eq.~\ref{positions} holds), where $V_{ij} = \langle v_i v_j \rangle$
are the elements of the velocity matrix,
\begin{equation}
\label{v}
C_{ij} = \langle (R_i - R^0_i)(R_j - R^0_j)\rangle
\end{equation}
are the elements of the displacement correlation matrix~\cite{henkes},
and angle brackets indicate averages over time. Vibrational
frequencies $\omega^s_k = \sqrt{s_k}$ can be obtained from the
eigenvalues of $S$. The binned versions of the density of vibrational
frequencies are given by $D(\omega^{s,d}) = [{\cal
N}(\omega^{s,d}+\Delta \omega^{s,d})-{\cal N}(\omega^{s,d})]/({\cal
N}(\infty)\Delta \omega^{s,d})$, where $N(\omega)$ is the number of
frequencies less than $\omega$. $D(\omega^d)$, $D(\omega^v)$, and
$D(\omega^s)$ are normalized so that $\int_0^\infty d\omega D(\omega)
= 1$.

{\it Results} In the harmonic vibrational response regime for MS
packings, the fluctuating particle positions are given by
Eq.~\ref{positions}, and $D(\omega^v)$ from the Fourier transform of
the velocity autocorrelation function is a set of $2N-2$ spikes at
frequencies $\omega^v_k$, which correspond to the eigenfrequencies
associated with the dynamical and displacement correlation matrices,
and become $\delta$-functions in the $T_0=0$ limit, $D(\omega^v) =
\sum_{k=1,2N-2} \delta(\omega^v - \omega^{d,s}_k)$.  In
Fig.~\ref{comparison}, we show $D(\omega^v)$ for MS packings at
$\Delta \phi > 0$, two $T_0$, and two system sizes.  For sufficiently
low $T_0$ in the ICS, the system displays harmonic response with
$\omega^v_k = \omega^{d,s}_k$. For larger $T_0$, the instantaneous
contact network deviates from that at $T_0=0$, and the vibrational
response becomes nonharmonic.  We find that $\omega_k^s < \omega_k^d$
(assuming the eigenvalues are sorted from smallest to largest),
$D(\omega^v)$ becomes a {\it continuous} spectrum, and none of the
densities of vibrational frequencies ($D(\omega^v)$, $D(\omega^s)$,
and $D(\omega^d)$) match. The same behavior is found for both $N=10$
and $128$.

\begin{figure*}
\includegraphics[width=1.02\textwidth]{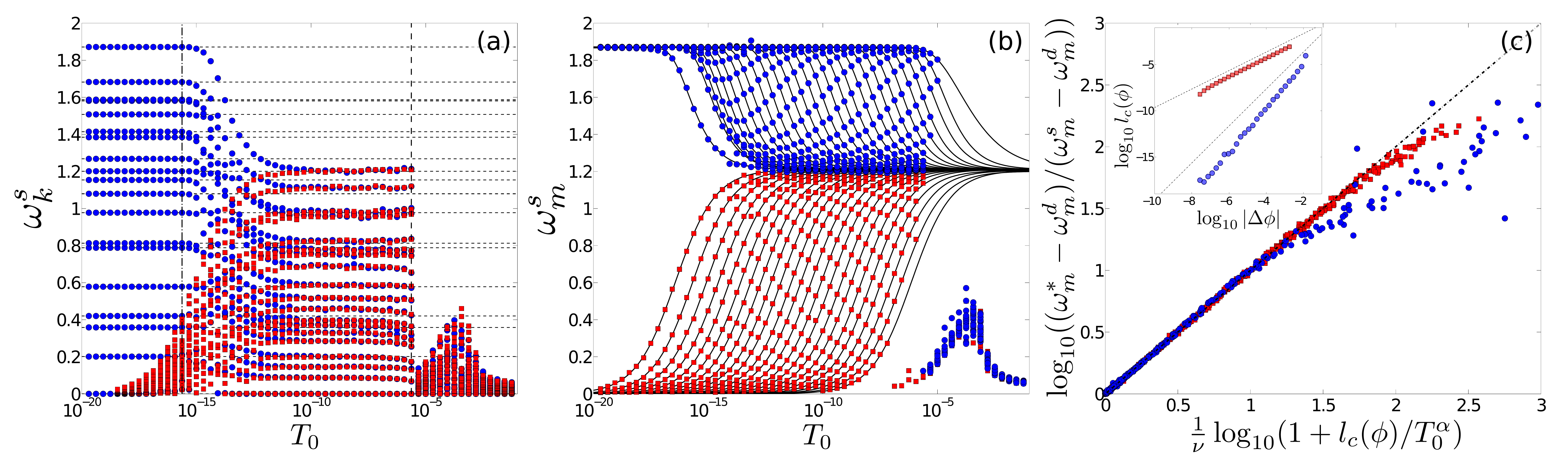}
\caption{(Color online) (a) The $2N-2$ nontrivial frequencies
$\omega^s_k$ associated with the displacement correlation matrix
versus $T_0$ for $\Delta \phi =10^{-6}$ (blue circles) and $-10^{-6}$
(red squares) for $N=10$. The vertical dot-dashed and dashed lines
indicate for $\Delta \phi=10^{-6}$ the $T_c$ where the contact network
first differs from that at $T_0=0$ and $T_{cb}$ where the
energy-minimized configurations differ from that at $T_0=0$,
respectively. The horizontal dashed lines show the dynamical matrix
eigenfrequencies. (b) The maximum frequency $\omega^s_m = \max_k
\omega^s_k$ versus $T_0$ for $28$ packing fraction deviations
logarithmically spaced in the range $10^{-7} \le |\Delta \phi| \le
10^{-2}$ above and below jamming. The solid lines are fits of
$\omega^s_m(T_0)$ to Eq.~\ref{logistic}. (c) Scaled $\omega^s_m$ versus
$T_0$. The inset shows $l_c$ versus $|\Delta \phi|$ above and below
jamming. The dashed and dotted lines have slope $1$ and $2$,
respectively.}
\vspace{-0.18in}
\label{spectrum}
\end{figure*}

In Fig.~\ref{spectrum} (a) we show the dependence of the frequencies
$\omega^s_k$ associated with the displacement correlation matrix
versus $T_0$ and $\Delta \phi$ for a $N=10$ MS packing above and below
jamming onset.  As shown in panel (a) for $\Delta \phi > 0$,
$\omega^s_k = \omega^d_k$ for all $k$ in the ICS when $T_0 <
T_c(\Delta \phi)$. For $T_0>T_c(\Delta \phi)$ and $\Delta \phi>0$, the
frequencies $\omega^s_k$ first decrease with $T_0$ and then each
reaches a $k$-dependent plateau value $\omega^*_k < \omega_k^d$ that
persists for more than six orders of magnitude (at $\Delta
\phi=10^{-6}$). The plateau ends abruptly after a particle
rearrangement occurs at $T_{cb}(\Delta \phi)$. Cage-breaking
rearrangements were identified by comparing the particle positions for
energy minimized configurations ${\vec R}^{\rm min}$ originally at
$T_0 >0$ to those for the MS packing ${\vec R}^0$ at $T_0=0$.  For
$T_0 < T_{cb}(\Delta \phi)$, the distribution $P(\Delta R)$ of
configurational distances $\Delta R = \sqrt{(2N)^{-1}\sum_{i=1}^{N}
[(x^{\rm min}_i - x^0_i)^2+(y^{\rm min}_i - y^0_i)^2] }$~\cite{gao2}
has a strong peak at small $\Delta R/\sigma_1 \sim 10^{-7}$ that
corresponds to the precision of the particle positions after energy
minimization.  For both $\Delta \phi > 0$ and $\Delta \phi<
0$, $P(\Delta R)$ is bimodal for $T \gtrsim T_{cb}(\Delta \phi)$ with
an additional well-separated peak at larger $\Delta R$ from
cage-breaking particle rearrangements. (See supplementary material.)

In contrast, for $\Delta \phi \le 0$, there is no harmonic vibrational
response regime.  We find that the frequencies $\omega_k^s$ increase
from zero at $T_0=0$ and reach the same $k$-dependent plateau values
$\omega^*_k$ as those for $\Delta \phi>0$.  These results point to a
robust set of frequencies (distinct from $\omega_k^d$) in the HCS
regime, where the contact network fluctuates, but the average contact
number remains constant near $\langle z\rangle/z_{\rm iso} \approx
0.5$.

Motivated by the behavior in the regime $\Delta \phi < 0$, for which
the frequencies $\omega^s_k \sim \sqrt{T_0}/l_c(\Delta \phi)$ scale
as the ratio of the velocity between interparticle collisions and
typical cage size and then reach a plateau $\omega_k^*$ at large
$T_0$, we propose the following scaling function:
\begin{equation}
\label{be}
\omega_k^s = \omega^*_k\left(1+\frac{l_c(\Delta \phi)}{\sqrt{T_0}}\right)^{-1},
\end{equation}
where $l_c$ is measured in units of $\sigma_1$. In Fig.~\ref{spectrum}
(b), we show least-squares fits of the maximum frequency associated
with the displacement correlation matrix $\omega^s_m$ versus $T_0$ to
Eq.~\ref{be} for $28$ logarithmically-spaced packing fraction
deviations below jamming in the range $10^{-7} \le |\Delta \phi| \le
10^{-2}$.  (Similar quality fits are found for all
lower frequencies.) Above jamming, $\omega^s_k$ interpolates between
$\omega_k^d$ at $T_0=0$ and $\omega^*_k$ at large $T_0$.  The
generalized logistic function,
\begin{equation}
\label{logistic}
\omega^s_k(T_0)=\omega^d_k+\frac{\omega^*_k-\omega^d_k}{(1+l_c(\Delta \phi)/T_0^{\alpha})^{\nu}},
\end{equation}
which allows variations of the slope of $\omega^s_k(T_0)$ in the
crossover region, is able to recapitulate $\omega^s_m$ both below and
above jamming (Fig.~\ref{spectrum} (b)). Below jamming, the
best fits give $\alpha = 0.5$ and $\nu=1$ ({\it i.e.}
Eq.~\ref{be}). Above jamming, we find $\alpha \sim
0.6$ and $\nu \approx 0.25 l_c^{-1/2}$ over a wide range of $\Delta
\phi$. In Fig.~\ref{spectrum} (c), we show that Eq.~\ref{logistic}
collapses $\omega^s_m$ from panel (b) with some deviation
at small $T_0$ for systems above jamming (caused by numerical accuracy
when $\omega_m^s \approx \omega^d_m$). In the inset to panel (c), we
show that the generalized cage size scales as $l_c \sim \Delta \phi$
below and $\sim (\Delta \phi)^{\lambda}$ with $\lambda \gtrsim 2$ above
jamming.

\begin{figure*}
\includegraphics[width=1.02\textwidth]{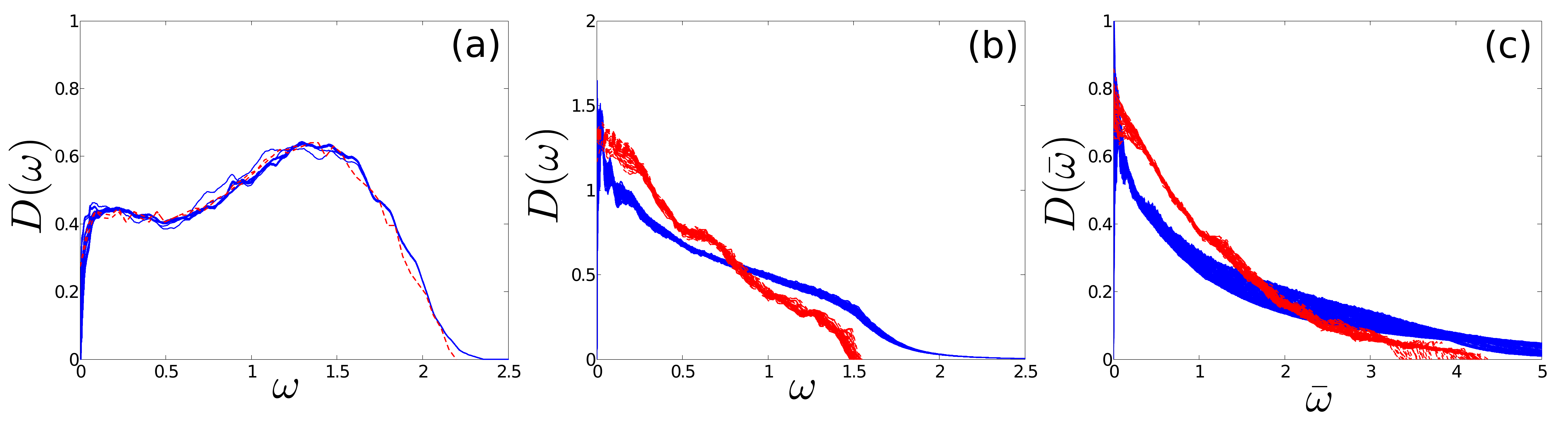}
\caption{(Color online) Binned density of vibrational frequencies
$D(\omega^v)$ (blue solid lines) and $D(\omega^s)$ (red dashed lines)
for more than $75$ sets of parameters in the (a) ICS and (b) HCS
regimes, as well as (c) hard-particle liquids for $N=128$. For HPL,
$\omega$ is normalized by the average frequency ${\overline \omega} = 
\omega/\langle \omega \rangle$.}
\label{dos}
\end{figure*}

We summarize our results for the measured vibrational response of MS
packings in the `phase diagram' in Fig.~\ref{phase_diagram}.  For
$\Delta \phi > \Delta \phi_c(T_0) \sim N^{\beta} \sqrt{T_0}/A$ (see
Ref.~\cite{schreck}), where $A \approx 0.5$ and $\beta \approx 0.85$,
the contact network does not change from that at $T_0=0$. In the ICS,
$\langle z\rangle = z_{\rm iso}$, the vibrational response is
harmonic, and the density of vibrational modes
$D(\omega^v)=D(\omega^s)=D(\omega^d)$ (Fig.~\ref{dos} (a)). Note that
the size of the ICS region decreases with increasing $N$. In
Fig.~\ref{phase_diagram}, we show that the midpoints of the crossovers
in $\langle z \rangle/z_{\rm iso}$ from $1$ to $0.5$, which define
$\Delta \phi_+^*(T_0)$, scale as $\sqrt{T_0}$. Assuming that the
effective particle diameter shrinks with $\sqrt{T_0}$~\cite{liu}, we
obtain an estimate for the shift in $\phi_J$ induced by thermal
fluctuations:
\begin{equation}
\label{shift} 
\Delta \phi_s(T_0) = \phi_J\left(\frac{1}{\left 
(1-\sqrt{2T_0}\right)^2}-1\right).
 \end{equation}
We find that near this boundary $\Delta \phi_s \sim \Delta \phi^*_+$
an extensive number of changes in the contact network from that at
$T_0=0$ has occurred. Similarly, for $\Delta \phi <0$, an extensive
number of `time-averaged' contacts has formed when $\Delta \phi > \Delta
\phi_-^*(T_0)$, which also scales as $\sqrt{T_0}$. In Fig.~\ref{dos}
(b), we show the density of vibrational frequencies for $75$ sets of
parameters in the HCS region $\Delta \phi > \Delta \phi_-^*(T_0)$,
$\Delta \phi < \Delta \phi_+^*(T_0)$, and $\Delta \phi > \Delta
\phi_{cb}(T_0)$.  We find that $D(\omega^s) \ne D(\omega^v)$, {\it
e.g.}  $D(\omega^s)$ possesses a stronger peak at low frequencies and
opposite curvature at high frequencies.  In contrast, $D(\omega^s)$
and $D(\omega^v)$ each separately can be collapsed for HPL by
scaling by the average frequency, but this does not mean that the
system is harmonic~\cite{berthier}.  In the transition regime, $\Delta
\phi_c(T_0) < \Delta \phi < \Delta \phi_+^*(T_0)$, $D(\omega)$ varies
continuously between that for the ICS and HCS in Fig.~\ref{dos} (a)
and (b).

{\it Conclusions} We emphasize that particulate systems over most of
the $\Delta \phi$ and $T_0$ plane near jamming display nonharmonic
vibrational response, {\it i.e.}  the density of vibrational
frequencies $D(\omega^v)$ from the Fourier transform of the velocity
autocorrelation function is continuous, not a set of discrete, sharp
peaks at the dynamical matrix eigenfrequencies $\omega_k^d$ and
differs from other measures of the vibrational response.  This implies
that when particulate systems are excited by a single mode
$\omega^d_k$, the response rapidly spreads to a continuous
spectrum of other modes.  This result has important consequences for
acoustic transmission~\cite{johnson} and thermal
transport~\cite{vitelli} in jammed solids.  However, there is a wide
swath of parameter space (corresponding to HCS) where the
frequencies associated with displacement correlation matrix do not
depend on $\Delta \phi$ or $T_0$.  $D(\omega^s)$ in this regime
differs from $D(\omega^d)$ for $\Delta \phi > 0$ at $T_0=0$ and from
$D(\omega^s)$ for hard-particle liquids.  These results underscore the
importance of measuring directly the vibrational response for MS packings
instead of inferring it from linear response.

We acknowledge support from NSF Grant No. CBET-0968013 (MS), DTRA
Grant No. 1-10-1-0021 (CO and TB), and Yale University (CS).
This work also benefited from the facilities and staff of the Yale
University Faculty of Arts and Sciences High Performance Computing
Center and the NSF (Grant No. CNS-0821132) that in part funded
acquisition of the computational facilities.

\end{document}